\documentclass[prb,twocolumn,amsmath,amssymb,
reprint,%
author-year,%
author-numerical,%
dvipdfmx,
]{revtex4-2}
\usepackage{graphicx}% Include figure files
\usepackage{bm}% bold math
\usepackage{color}
\begin{document}

%\preprint{APS/123-QED}
\title{Two-dimensional structures formed by pair potentials:\\
An analytical and numerical study}
\author{Shota Ono}
\email{shotaono@muroran-it.ac.jp}
\affiliation{Department of Sciences and Informatics, Muroran Institute of Technology, Muroran 050-8585, Japan}

\begin{abstract}
Understanding the thickness dependence of the excess energy of thin films and identifying the deviation from the scaling law that holds for thicker films are crucial for predicting the formation of two-dimensional (2D) materials derived from nonlayered structures, commonly referred to as non-van der Waals (non-vdW) 2D materials. Here, we analytically study the excess energy of thin films as a function of the number of layers $N$, assuming pair potentials. We show that the $1/N$ scaling law of the excess energy breaks down in the monolayer limit when the interatomic interactions are long-ranged. The analytical results are confirmed by numerical simulations using the Lennard-Jones and Morse potentials. The present work suggests that non-vdW 2D structures can be formed by pair potentials alone. 
\end{abstract}
%\pacs{}

\maketitle

%%%%%%%%%%%%%%%%%%%%%%%%%%%%%%%%%%%%%%%%%%%%%%
\section{Introduction}
Non-van der Waals (non-vdW) two-dimensional (2D) materials are monolayers derived from nonlayered bulk structures. Covalently bonded group-IV elements are known to form buckled honeycomb structures \cite{vogt,fleurence,davila,molle2025}. Potential structures of elemental 2D metals have been investigated using first-principles approaches \cite{koskinen2018,ono2020,koskinen2024}. Plumbene (2D Pb) and goldene (2D Au) have been synthesized \cite{yuhara,kashiwaya}. However, experimental realization of 2D metallic monolayers remains rare, although metallic thin films or membranes with sub-nanometer thicknesses have also been reported \cite{ren2025,zhao2025}. Hypothetical non-vdW 2D compounds have been predicted through first-principles calculations, with several descriptors, such as the anisotropy of force constants \cite{jia,bagheri2023,bagheri2026} and the exfoliation energy \cite{friedrich2022,friedrich2026,zhong2025}, employed to screen the materials space. However, many predicted 2D materials remain unrealized, highlighting the gap between theoretical predictions and experimental synthesis \cite{x2db,ono2026_jpcm}. 

Recently, the author studied the thickness dependence of the excess energy of thin films relative to the bulk energy and demonstrated that the breakdown of the $1/N$ scaling law is a useful indicator for predicting non-vdW 2D materials such as silicene, germanene, plumbene, and goldene, where $N$ is the number of layers \cite{ono2025}. While first-principles approaches are useful for accurately predicting the excess energies of realistic 2D materials \cite{ono2026}, developing an analytical model to capture the underlying physics behind the $1/N$ scaling law would be useful. 

In this paper, we analytically study the excess energy of thin films formed by pair potentials. The behavior of the excess energy at $N=1$ depends on the range of the interatomic interactions. For short-range interaction, the $1/N$ scaling law holds even in the monolayer limit. On the other hand, if the interaction becomes long-ranged, the breakdown of the $1/N$ scaling law is observed for small $N$. This is confirmed by numerical calculations using the Lennard-Jones (LJ) and Morse potentials. Thin films formed by pair potentials serve as a minimal model for the formation of non-vdW 2D structures. 

%%%%%%%%%%%%%%%%%%%%%%%%%%%%%%%%%%%%%%%%%%%%%%
\section{Theory}
\label{sec:theory}
The aim of this section is to formulate the excess energy of $N$ layer thin films. To analytically derive the thickness dependence of the excess energy, we assume pair potentials for the interatomic interactions.  

\subsection{Total energy of $N$ layer thin films}
Let us consider a monatomic crystal in which each particle interacts with other particles through a pair potential $v(R)$, where $R$ is the interatomic distance. The crystal exhibits a close-packed structure consisting of triangular lattices stacked along the surface normal. We construct an $N$-layer thin film derived from the close-packed surface and examine the $N$-dependence of the total energy below. In our analytical study, geometry relaxation in thin films is not considered. 

For $N=1$, i.e., a monolayer triangular lattice, the total energy is a sum of the two-body interaction energy between atoms. This is expressed by
\begin{eqnarray}
 E_1 = \frac{1}{2} \sum_\mu \sum_{\nu (\ne \mu)} v (R_{\mu\nu}^{1,1}), 
 \label{eq:e0}
\end{eqnarray}
where the factor $1/2$ accounts for the double counting. We have defined the interatomic distance between the atom $\mu$ in the layer $n$ and the atom $\nu$ in the layer $m$ as
\begin{eqnarray}
 R_{\mu\nu}^{n,m} = \sqrt{(x_{\mu}^{n} - x_{\nu}^{m})^2 
 + (y_{\mu}^{n} - y_{\nu}^{m})^2 + (z_{\mu}^{n} - z_{\nu}^{m})^2},
\end{eqnarray}
where $x_{\mu}^{n},y_{\mu}^{n}$, and $z_{\mu}^{n}$ are the Cartesian coordinates of the atom $\mu$ in the layer $n$. For the adjacent interlayer distance of $d$, $z_{\mu}^{n} - z_{\nu}^{m}$ may be expressed by $\vert n - m \vert d$. 

For $N$ thin films, the total energy is a sum of the intralayer energy for each layer and the interlayer energy between different layers, which is given by
\begin{eqnarray}
 E_N = N E_1 + \frac{1}{2} \sum_{n=1}^{N} \sum_{m\ne n}^{N} V(n,m),
\end{eqnarray}
where the summation in the second term is restricted to $n\ne m$. The interlayer energy between the $n$th and $m$th layers is defined as
\begin{eqnarray}
V(n,m) = \sum_{\mu}^{(n)} \sum_{\nu}^{(m)} v( R_{\mu\nu}^{n,m}),
\end{eqnarray}
where the sums over $\mu$ and $\nu$ runs over the atoms in the $n$th and $m$th layers, respectively. By definition, 
\begin{eqnarray}
& & V(n,m) = V(m,n),
\\
& & V(n,n+l) = V(m,m+l)
\label{eq:Vnm}
\end{eqnarray}
for an integer $l$ satisfying $1\le n (m)+l \le N$. For the bulk limit, the total energy per layer is defined as
\begin{eqnarray}
E_{\rm bulk} &=& \lim_{N\rightarrow \infty} \frac{E_N}{N} 
= E_1 + E_{\rm int},
\end{eqnarray}
where the interaction energy of bulk is expressed by
\begin{eqnarray}
E_{\rm int} = \lim_{N\rightarrow \infty} \frac{1}{2N} \sum_{n=1}^{N} \sum_{m\ne n}^{N} V(n,m).
\end{eqnarray}

%%%%%%%%%%%%%%%%%%%%%%%%%%%%%%%%%%%%%%%%%%%%%%
\subsection{Excess energy of $N$ layer thin films}
We measure the excess energy of thin films from the bulk energy as in Ref.~\cite{ono2025}. This is defined as 
\begin{eqnarray}
\Delta E(N) &=& \frac{E_N}{N} - E_{\rm bulk}, 
\end{eqnarray}
which is expressed as
\begin{eqnarray}
\Delta E(N) 
&=& \frac{1}{2N} \sum_{n=1}^{N} \sum_{m\ne n}^{N} V(n,m)- E_{\rm int}.
\end{eqnarray}
This is equal to the difference between the interlayer interaction energy per layer for the $N$-layered thin film and that in the bulk. 

%%%%%%%%%%%%%%%%%%%%%%%%%%%%%%%%%%%%%%%%%%%%%%
\subsubsection{Empirical rule for realizing non-vdW 2D materials}
In the previous study \cite{ono2025}, we have shown that the $\Delta E(N)$ exhibits the $1/N$ scaling law for large $N$ within a phenomenological discussion, where the thin film energy is assumed to be $E_N \simeq 2E_{\rm surf} + (N-2)E_{\rm bulk}$ with $E_{\rm surf}$ being the energy of the surface layer. In addition, by performing first-principles calculations, we have demonstrated that specific elements (e.g., silicon and gold) exhibit the downward deviation from the $1/N$ scaling law in the monolayer limit. To evaluate the magnitude of the deviation, we have calculated the following quantity \cite{ono2025}
\begin{eqnarray}
 p_1 = \frac{d\ln \Delta E(N)}{d\ln N}\Big\vert_{N=1} \simeq \frac{\ln \Delta E(2) - \ln \Delta E(1)}{\ln 2}.
\label{eq:p1}
\end{eqnarray}
If $\Delta E(N)\propto 1/N$ even at $N=1$, $p_1 = -1$. On the other hand, if $\Delta E(N)$ exhibits the downward deviation, $p_1 > -1$. For example, silicene and goldene show $p_1 = -0.38$ and $-0.58$, respectively.  

Below, we provide an analytical discussion that the long-range interlayer interaction energy $V(n,m)$ is important for obtaining $p_1 > -1$. If $V(n,m)$ is short-ranged, the $1/N$ scaling law holds even at $N=1$, but if $V(n,m)$ is long-ranged, the deviation from the $1/N$ scaling law can appear. In Sec.~\ref{sec:numerical}, we will demonstrate how the range of the interatomic interaction $v(R)$ modulates that of $V(n,m)$ by assuming the LJ and Morse potentials.  

%%%%%%%%%%%%%%%%%%%%%%%%%%%%%%%%%%%%%%%%%%%%%%
\subsubsection{Short-range interaction}
\label{sec:short-range}
We first consider that $V(n,m)$ is short-ranged enough to assume that the interaction energies between only adjacent layers are dominant. In such thin films, the contribution to the excess energy from the surface layers (i.e., layers $n=1$ and $N$) is larger than that from the inner layers (i.e., $n=2, 3, \cdots, N-1$). Therefore, $(N-2)/N$ of the interlayer interaction energy of bulk ($E_{\rm int}$) is cancelled by the excess energy in the inner layers of the thin film. We then obtain
\begin{eqnarray}
\Delta E(N) 
%&=& 
%\frac{1}{2N} \left[ \sum_{m\ne 1}^{N} V(1,m) + \cdots + \sum_{m\ne N}^{N} V(N,m) \right]
%\nonumber\\
%&-& \left( \frac{N-2}{N} + \frac{2}{N} \right) E_{\rm int} 
%\nonumber\\
&\simeq & \frac{1}{2N} \left[ \sum_{m\ne 1}^{N} V(1,m) + \sum_{m\ne N}^{N} V(N,m) \right]
- \frac{2}{N} E_{\rm int} 
\nonumber\\
&=  & \frac{1}{N} \sum_{m\ne 1}^{N} V(1,m)
- \frac{2}{N} E_{\rm int} 
\end{eqnarray}
In the last line, we used the fact that the contribution from $n=1$ (bottom layer) is the same as that from $n=N$ (top layer). The partial cancellation of the excess energy between thin films and bulk holds when 
\begin{eqnarray}
E_{\rm int} \simeq \frac{1}{2} \sum_{m\ne 2}^{N} V(2,m).
\label{eq:app1}
\end{eqnarray}
We have assumed that the contributions from $n=2,3, \cdots, N-1$ are the same. Thus, we obtain 
\begin{eqnarray}
\Delta E(N) \simeq \frac{1}{N} \sum_{m\ne 1}^{N} V(1,m) 
- \frac{1}{N} \sum_{m\ne 2}^{N} V(2,m)
\label{eq:DE1}
\end{eqnarray}
%This is expressed by
%\begin{eqnarray}
%\Delta E(N) &\simeq& \frac{1}{N} \Large[ 
%E_{\rm inter}^{(1,2)} + E_{\rm inter}^{(1,3)} + \cdots + E_{\rm inter}^{(1,N)}
%\nonumber\\
%&-& E_{\rm inter}^{(2,1)} - E_{\rm inter}^{(2,3)} - \cdots - E_{\rm inter}^{(2,N)} \Large].
%\end{eqnarray}
With Eq.~(\ref{eq:Vnm}), several terms are cancelled out in Eq.~(\ref{eq:DE1}). Finally, we obtain 
\begin{eqnarray}
\Delta E(N) &\simeq& \frac{1}{N} \left[ 
 V(1,N) - V(2,1) \right] 
 \nonumber\\
 &\simeq& - \frac{1}{N} V(2,1),
\end{eqnarray}
which is proportional to $1/N$. We assumed $\vert V(1,N) \vert \ll \vert V(2,1) \vert$ and $V(2,1)<0$. 
When this is extrapolated to $N=1$, one obtains 
\begin{eqnarray}
\Delta E_{1}^{\rm ext} = \lim_{N\rightarrow 1} \Delta E(N) = - V(2,1). 
\label{eq:DE1ext}
\end{eqnarray}

On the other hand, at $N=1$, the total energy is $E_1$ exactly. Therefore, with Eq.~(\ref{eq:app1}) again, one obtains
\begin{eqnarray}
\Delta E(1) &=& E_1 - (E_1 + E_{\rm int}) \simeq 
- \frac{1}{2} \sum_{m\ne 2}^{N} V(2,m).
\label{eq:DE1short}
%\nonumber\\
%&=&  - \frac{1}{2} \left[ E_{\rm inter}^{(2,1)} + E_{\rm inter}^{(2,3)} + \cdots \right]  
\end{eqnarray}
As the interlayer energies between adjacent layers are dominant (i.e., $\vert V(2,1)\vert =\vert V(2,3)\vert \gg \vert V(2,m)\vert$ for $m\ge 4$), we obtain
\begin{eqnarray}
\Delta E(1) \simeq -  V(2,1),
\end{eqnarray}
which is exactly equal to Eq.~(\ref{eq:DE1ext}) and thus follows the $1/N$ scaling law. 

%%%%%%%%%%%%%%%%%%%%%%%%%%%%%%%%%%%%%%%%%%%%%%
\subsubsection{Long-range interaction}
\label{sec:long-range}
The analysis of the excess energy described in Sec.~\ref{sec:short-range} can be extended to the case for the long-range interlayer interactions. We now assume that the contribution to the excess energy from the $l$ surface layers (i.e., $n=1, 2,\cdots,l $ and $N-l+1, N-l+2,\cdots, N$) is larger than that from the inner layers (i.e., $n=l+1, l+2, \cdots N-l$). Then, $(N-2l)/N$ of the interlayer interaction energy of bulk is cancelled by the excess energy in the inner layers of the thin film. The inequality $N\ge 2l+1$ is satisfied when more than one inner layer exists. In addition, we assume that the $n$th layer from the top has the same energy as the $n$th layer from the bottom for $n\le l$. With these approximations, we obtain 
\begin{eqnarray}
\Delta E(N) 
%&=& \frac{1}{2N} \left[ \sum_{m\ne 1}^{N} E_{\rm inter}^{(1,m)}
%+ \cdots + \sum_{m\ne N}^{N} E_{\rm inter}^{(N,m)} \right]
%\nonumber\\
%&-& \left( \frac{N-2l}{N} + \frac{2l}{N} \right) E_{\rm int} 
%\nonumber\\
%&\simeq & \frac{1}{2N} \Big[ \sum_{m\ne 1}^{N} E_{\rm inter}^{(1,m)}
%+\cdots + \sum_{m\ne l}^{N} E_{\rm inter}^{(l,m)} 
%\nonumber\\
%&+& \sum_{m\ne N-l+1}^{N} E_{\rm inter}^{(N-l+1,m)} 
%+ \cdots + \sum_{m\ne N}^{N} E_{\rm inter}^{(N,m)} \Big]
%- \frac{2l}{N} E_{\rm int} 
%\nonumber\\
&\simeq & \frac{1}{N} \left[ \sum_{m\ne 1}^{N} V(1,m)
+\cdots + \sum_{m\ne l}^{N} V(l,m) \right]
\nonumber\\
&-& \frac{2l}{N} E_{\rm int}
\label{eq:DE_long}
\end{eqnarray}
under the condition of the excess energy cancellation 
\begin{eqnarray}
E_{\rm int} \simeq \frac{1}{2} \sum_{m\ne l+1}^{N} V(l+1,m).
\end{eqnarray}
%Thus, we obtain 
%\begin{eqnarray}
%\Delta E(N) &\simeq& \frac{1}{N} \Big[ \sum_{m\ne 1}^{N} E_{\rm inter}^{(1,m)}
%+\cdots + \sum_{m\ne l}^{N} E_{\rm inter}^{(l,m)} \Big]
%\nonumber\\
%&-& \frac{l}{N} \sum_{m\ne l+1}^{N} E_{\rm inter}^{(l+1,m)}
%\end{eqnarray}
%this is simplified into 
%\begin{eqnarray}
%\Delta E(N) &\simeq& \frac{1}{N} \Big[ 
% E_{\rm inter}^{(1,N-l+1)} + \cdots + E_{\rm inter}^{(1,N)} 
%\nonumber\\
%&+&  E_{\rm inter}^{(2,N-l+2)} + \cdots + E_{\rm inter}^{(2,N)} 
%\nonumber\\
%&+& \cdots + E_{\rm inter}^{(l,N)} 
%\nonumber\\
%&+& E_{\rm inter}^{(2,1)} + E_{\rm inter}^{(3,1)} + E_{\rm inter}^{(3,2)}
%\nonumber\\
%&+& \cdots + E_{\rm inter}^{(l,1)} + \cdots + E_{\rm inter}^{(l,l-1)} \Big]  
%\nonumber\\
%&-& 
% \frac{l}{N} \Big[ E_{\rm inter}^{(l+1,1)} + \cdots E_{\rm inter}^{(l+1,l)}\Big]
%\end{eqnarray}
As in Sec.~\ref{sec:short-range}, we ignore the interlayer interaction energy for which the interlayer distance satisfies $z\ge (l+1)d$, where $d$ is the interlayer spacing. With a straightforward calculation, one obtains  
\begin{eqnarray}
\Delta E(N) &\simeq& - \frac{1}{N} \sum_{k=1}^{l} 
k V(l+1,l-k+1),
\end{eqnarray}
which is proportional to $1/N$. At $N=1$, this is extrapolated to
\begin{eqnarray}
\Delta E_{1}^{\rm ext} = -\sum_{k=1}^{l} 
k V(l+1,l-k+1).
\label{eq:DE2ext}
\end{eqnarray}

On the other hand, Eq.~(\ref{eq:DE1short}) derived for the short-range interactions can be generalized to
\begin{eqnarray}
\Delta E(1) &\simeq& 
- \frac{1}{2} \sum_{m\ne l+1}^{N} V(l+1,m)
\nonumber\\
&\simeq&  - \left[ V(l+1,1) + \cdots + V(l+1,l)\right].
\end{eqnarray}
Note that the following difference 
\begin{eqnarray}
\Delta E(1) - \Delta E_{1}^{\rm ext} 
%&=& - \left[ E_{\rm inter}^{(l+1,1)} + \cdots + E_{\rm inter}^{(l+1,l)}\right]  
%\nonumber\\
%&+&  \left[ l E_{\rm inter}^{(l+1,1)} + \cdots + 2 E_{\rm inter}^{(l+1,l-1)} + E_{\rm inter}^{(l+1,l)} \right]
%\nonumber\\
%&=& (l-1) V(l+1,1) + \cdots + V(l+1,l-1)
&=& \sum_{k=1}^{l-1} k V(l+1,l-k) 
\label{eq:diff}
\end{eqnarray}
is negative because $V(n,m)<0$. Therefore, the excess energy at $N=1$ is smaller than the theoretical curve of the $1/N$ scaling law, indicating the downward deviation from the $1/N$ scaling law. 

Note that the expressions for $\Delta E(N), E_{\rm int}$, and $\Delta E_{1}^{\rm ext}$ given in Sec.~\ref{sec:short-range} can be recovered from those derived in this section by assuming $l=1$. For $l\ge 2$, Eq.~(\ref{eq:diff}) shows that $V(l+1,l-1)$, which is the next-nearest interlayer interaction energy, begins to contribute to the deviation from the $1/N$ scaling law.

%%%%%%%%%%%%%%%%%%%%%%%%%%%%%%%%%%%%%%%%%%%%%%
\section{Numerical simulation}
\label{sec:numerical}
To provide a concrete example illustrating the theory developed in Sec.~\ref{sec:theory}, we study the crystals formed by the LJ and Morse potentials that are implemented in the Atomic Simulation Environment (ASE) \cite{ase}. Below, we consider $N$-layer thin films derived from the (111) surface of a face-centered cubic structure that exhibits the ABC stacking. Thin films with the AB stacking, derived from a hexagonal close-packed structure, are almost degenerate in energy with their ABC stacking counterparts. The structural optimization was performed using the Broyden-Fletcher-Goldfarb-Shanno (BFGS) method \cite{numerical_recipe}. 

The LJ potential is defined as
\begin{eqnarray}
 v_{\rm LJ}(R) = 4\varepsilon \left[ \left( \frac{\sigma}{R} \right)^{12} 
 - \left( \frac{\sigma}{R} \right)^{6} \right] C(R),
\end{eqnarray}
where $\varepsilon$ and $\sigma$ are the parameters that determine the energy and length scales, respectively. The cutoff function is assumed to be
\begin{eqnarray}
 C (R) = \frac{(R_{c}^{2} - R^{2})^2 (R_{c}^{2} +2 R^{2} - 3R_{o}^{2}) }{ (R_{c}^{2} - R_{o}^{2})^3 },
\end{eqnarray}
where $R_c$ and $R_o = 0.66 R_{c}$ are the cutoff radii, which ensure continuity of the energy and forces \cite{jaxmd}. In the present work, we tuned $R_c$ to change the range of the pair interactions, and simultaneously, the range of the interlayer interactions.  

The Morse potential is defined as
\begin{eqnarray}
 v_{\rm M}(R) = \varepsilon \left[ 
 e^{-2\rho_0 \left(s - 1\right)} - 2e^{-\rho_0 (s - 1)}
 \right] S_2(R)
\end{eqnarray}
with a dimensionless distance $s = R/r_0$. $\varepsilon$ and $r_0$ are the parameters that determine the energy and length scales, respectively. $S_2(R)$ is a sigmoid-like function that smoothly interpolates between two points, $(R,S_2)=(1.9r_0,1)$ and $(2.7r_0,0)$, where the default values for the cutoff range have been assumed \cite{ase}. We tuned the exponential prefactor $\rho_0$. A small $\rho_0$ gives a moderate change of the pair potential, yielding long-range interlayer interactions.  

The interlayer interaction energy $V(1,m)$ can be determined recursively from the total energies of thin films up to the $m$ layers and the interaction energies up to $V(1,m-1)$. For $m=2$,
\begin{eqnarray}
 V(1,2) &=& E_2 - 2E_1,
\end{eqnarray}
and for $m\ge 3$, 
\begin{eqnarray}
 V(1,m) &=& E_m - mE_1
 - \sum_{k=2}^{m-1} (m-k+1) V(1,k). \ \ \ 
\end{eqnarray}

%%%%%%%%%%%%%%%%%
\begin{figure}
\center\includegraphics[scale=0.45]{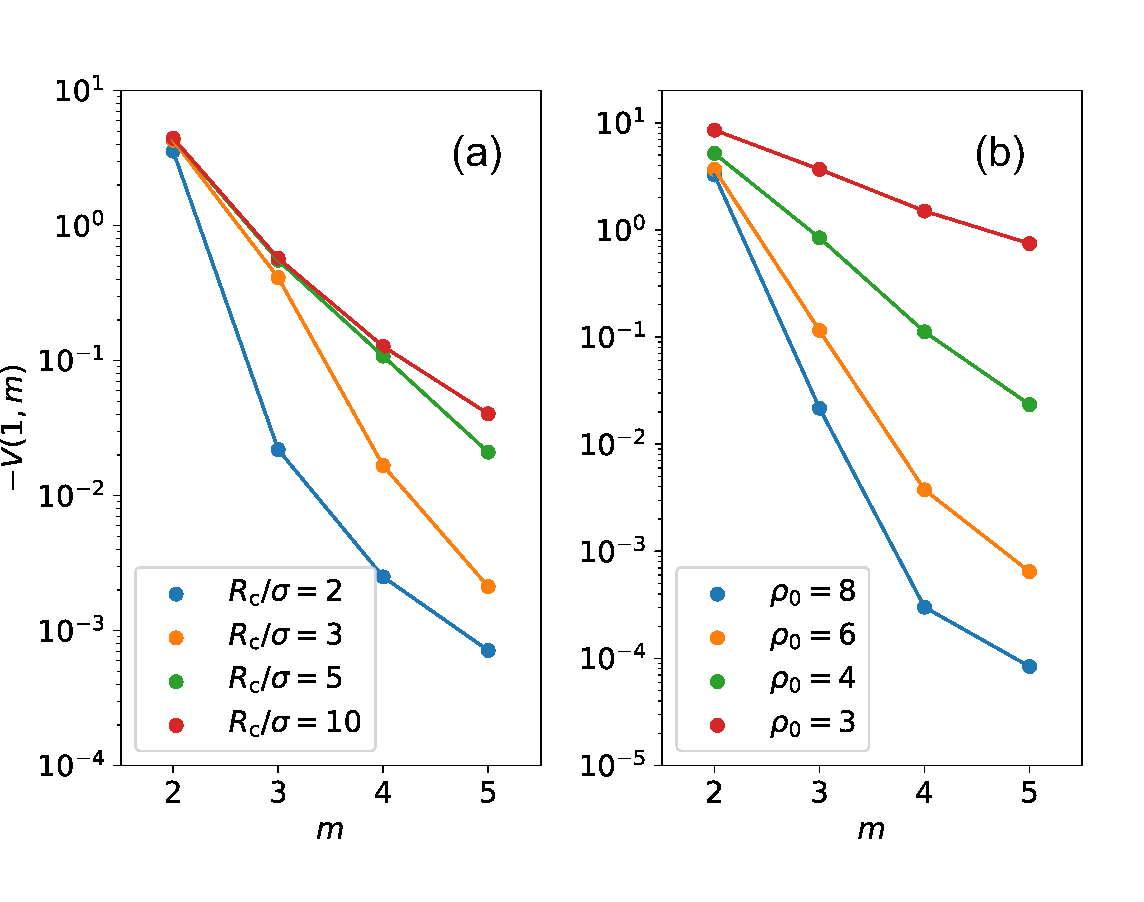}
\caption{Negative interlayer interaction energy $-V(1,m)$ for (a) the LJ and (b) the Morse thin films, where the cutoff radii $R_{c}$ and the exponential prefactors $\rho_0$ are tuned, respectively. } \label{fig1} 
\end{figure}
%%%%%%%%%%%%%%%%%

%%%%%%%%%%%%%%%%%
\begin{figure}
\center\includegraphics[scale=0.43]{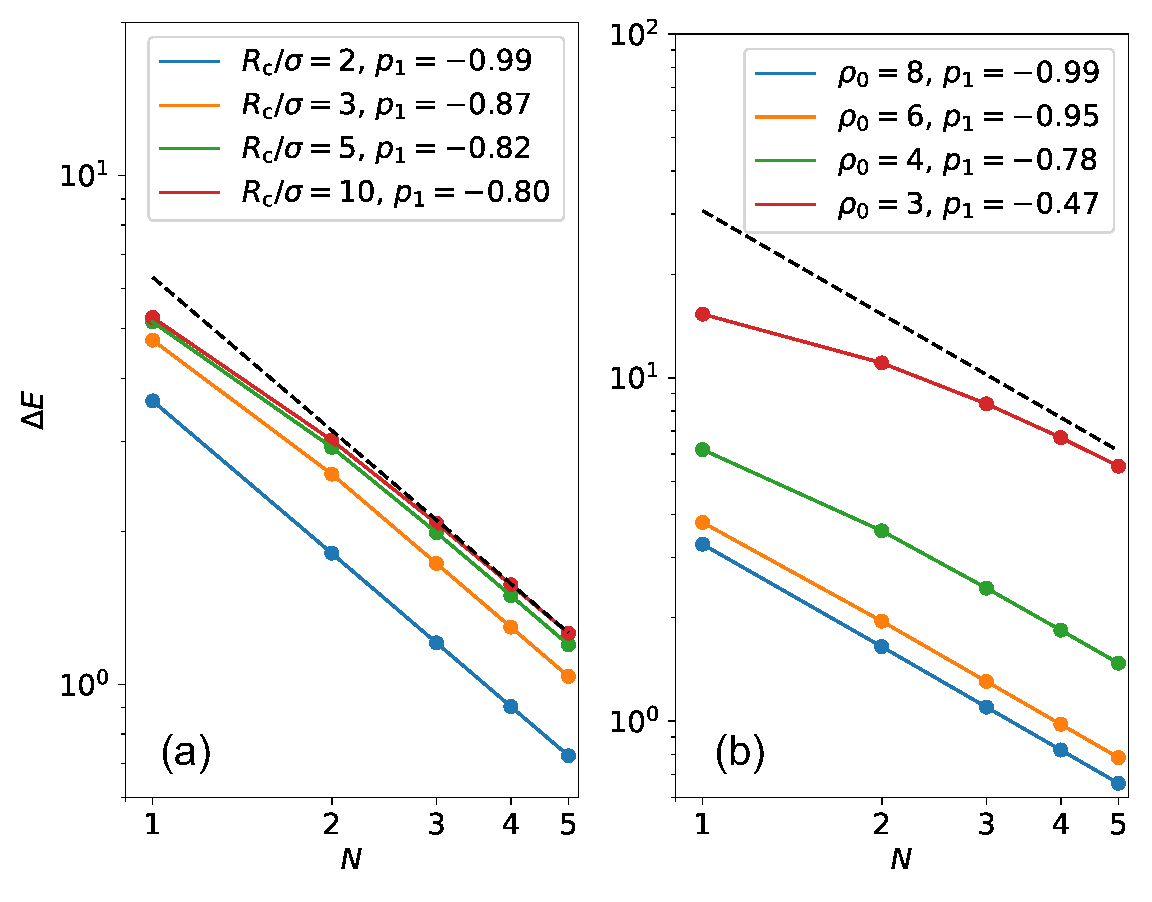}
\caption{The $N$-dependence of the excess energy of thin films formed by (a) the LJ potentials parametrized by the cutoff radii $R_{c}$ and (b) the Morse potentials parametrized by the exponential prefactors $\rho_0$. The downward deviation from the $1/N$ scaling law is observed when $R_{\rm c}$ is large and $\rho_0$ is small. The value of $p_1$ in Eq.~(\ref{eq:p1}) is indicated. } \label{fig2} 
\end{figure}
%%%%%%%%%%%%%%%%%

Figure \ref{fig1} shows how two parameters, $R_c$ for the LJ potential and $\rho_0$ for the Morse potential, affect the range of the interlayer interactions $V(1,m)$. When $R_{\rm c} = 2\sigma$ (LJ) and $\rho_0=8$ (Morse), $\vert V(1,2) \vert$ is more than a hundred times larger than $\vert V(1,m) \vert$ with $m\ge 3$. This indicates that the interatomic interaction is sufficiently short-ranged, so that the interlayer interaction is also short-ranged. As $R_{\rm c}$ increases or $\rho_0$ decreases, the magnitude of $\vert V(1,m) \vert$ with $m\ge 3$ becomes large, and the interaction range becomes long-ranged. For the LJ thin films, the curves of $R_{\rm c} = 5\sigma$ and $10\sigma$ almost overlap. On the other hand, the range of interlayer interaction is further increased as $\rho_0$ decreases. No convergence was obtained when $\rho_0 \le 2$ for the Morse potential. 

Figure \ref{fig2} shows $\Delta E(N)$ of $N$-layered thin films formed by (a) the LJ and (b) the Morse potentials. For short-range interaction, i.e., $R_{\rm c} = 2\sigma$ (LJ) and $\rho_0=8$ (Morse), the excess energy follows the $1/N$ scaling law even at $N=1$, yielding $p_1\simeq -1$. As $R_{\rm c}$ increases or $\rho_0$ decreases, the interaction range becomes long-ranged, yielding the downward deviation from the $1/N$ scaling law. When $R_{\rm c}=10\sigma$ (LJ) and $\rho_0=3$ (Morse), $p_1=-0.8$ and $-0.47$ are obtained, respectively. Thus, the numerical results confirm the analytical result demonstrated in Sec.~\ref{sec:theory} that long-range interlayer interactions are relevant for the breakdown of the $1/N$ scaling law. 

Even without geometry relaxation, the excess energy exhibits the downward deviation from the $1/N$ scaling law. We obtain $p_1 = -0.8$ and $-0.31$ for $R_{\rm c}=10\sigma$ (LJ) and $\rho_0=3$ (Morse), respectively. These results also support the assumption introduced in Sec.~\ref{sec:theory}, where geometry relaxation has a minor impact on the scaling law.  

\section{Conclusion}
The downward deviation of the excess energy from the $1/N$ scaling law is a useful descriptor for identifying non-vdW 2D materials \cite{ono2025}. To construct a model system exhibiting a breakdown of the $1/N$ scaling law, we have studied the thickness dependence of the excess energy of thin films formed by pair potentials. Our analytical results show that the $1/N$ scaling law of the excess energy breaks down in the monolayer limit when the interatomic interactions are long-ranged. This has been confirmed by numerical simulations using the LJ and Morse potentials. 

The present work suggests that non-vdW 2D structures can be formed by pair potentials alone, provided that the interactions are long-ranged. The effects of many-body interactions (e.g., chemical bonds) beyond the pair potential approximation should be investigated using density-functional theory. The effects of substrates, encapsulation, and other environmental factors (e.g., strains) on the formation of non-vdW 2D structures are beyond the scope of the present work. 

\begin{acknowledgments}
This work was supported by JSPS KAKENHI (Grant No. 24K01142). 
\end{acknowledgments}

%===================================================================%
%   References
%===================================================================%

%\bibliography{refs}

\end{document}